\documentstyle[11pt]{article}
\begin{document}
\baselineskip=1.0truecm
\begin{titlepage}
\title{
Is the Strong Anthropic Principle Too Weak?}
\author{A. Feoli  and S. Rampone \\ Dipartimento di Scienze Fisiche
``E.R.Caianiello'' \\ Universit\`a di Salerno \\
I-84081  Baronissi (Salerno)\\
Italy}
\date{\empty}
\maketitle
\begin{abstract}
We discuss the Carter's formula  about the 
mankind evolution probability
following the derivation
proposed by Barrow and Tipler.
We stress the relation between 
the existence of billions of galaxies and  the
evolution of  at least one intelligent life, 
whose living time is not trivial,
all over the Universe.
We show that the existence probability and the lifetime of a civilization
depend not only on the evolutionary critical steps, but also 
on the number of places where the life can arise. 
In the light of  these results, 
we propose a stronger
version of Anthropic Principle.
\end{abstract}

\end{titlepage}

{\rightline {\em Entia non sunt multiplicanda praeter necessitatem}}

{\rightline {\em W. Occam}}

\section{Introduction}
In a seminal paper~\cite{1}, Brandon Carter proposed two versions of the 
Anthropic Principle $(AP)$. The weak interpretation, against the ``dogma''
of the Copernican principle, takes into account ``the fact that our
location
in the Universe is necessarily privileged to the extent of being compatible
with our existence as observers''. Furthermore he
called Strong Anthropic Principle {\em (SAP)} 
the statement {\em ``The Universe (and hence the fundamental 
parameter on which it depends) must be such as to admit the creation of
observers within it at some stage''}.
Later on many versions and
interpretations of the AP have been proposed. 
A collection of the anthropic arguments is contained
in the Barrow and Tipler's book~\cite{2} where four different statements of
AP are defined: Strong {\em (SAP)}, Weak {\em (WAP)},
Participatory {\em (PAP)}, and Final {\em (FAP)} Anthropic Principle. 
While some of these statements appear to have teleological overtones,
AP proper sense is that our existence, as intelligent life form evolved on a
earth-like 
planet, is a matter of fact, and Universe laws can not contradict this
fact.

Starting from the AP, 
Carter~\cite{3} connected the number $n$ of very
improbable steps in the Homo Sapiens evolution to the existence 
length $t_0$ of a biosphere, and to the evolution time $t_e$ required to
produce an intelligent species on an earth-like planet. 
Carter's estimation, 
discussed in Section 2, of how long a biosphere
will continue to exist after an intelligent life evolution was 
$t_0 - t_e = t_0/(n+1)$. By the  experimental
evidence of our own evolution completed in a time  $ t_e \simeq 0.4 t_0 $
Carter was forced to conclude that there are
at most
two critical steps, even if  he 
``had previously inclined to think that the appropriate value of $n$ [...]
was likely to be very large".

Later Barrow and Tipler~\cite{4} estimated in fact a much larger $n$, 
and they used this argument just to exclude 
the existence of extraterrestrial human like beings.  
Unfortunately this result holds also for the Homo Sapiens. It gives 
the enormous improbability 
of the evolution of intelligent life in general, and on Earth
in particular.

Since Barrow and Tipler's book, the scientific and philosophical
debate about AP has been going on  with some criticisms and some
enthusiastic
supporters~\cite{5}. 

For example Rosen~\cite{6} shows ``the conviction that AP is among the
most important fundamental principles around, even [...]
the most basic principle we have.
What physical phenomenon in the whole wide world are we most sure of,
have least doubts about, have the most confidence in? The answer is our
own existence. Thus the most physical explanation is one based on our
own existence, and that is what is so special about AP".

On the contrary Hawking~\cite{7} thinks that ``it runs against the tide
of the whole history of science. [...] The Earth is a medium-size
planet [...] in the outer suburbs of an ordinary spiral galaxy, which is
itself one of about a million million galaxies in the observable Universe. 
Yet the SAP would claim that this whole vast construction exists simply for
our
sake. Our Solar System is certainly a 
prerequisite for our existence, and one might extend this to the whole of
our galaxy to allow for an earlier generation of stars that created the
heavier
elements. But there does not seem to be any need for all those galaxies,
nor for
the Universe to be so uniform and similar in every direction on the large
scale.'' 

This is not an objection to the SAP really, but to some teleological
arguments
related to SAP. Since its appearance, SAP has been forced to have a
teleological
meaning. For example, Press explicitly warns the readers of Barrow and
Tipler's
book that the authors want to convince them ``of an astounding claim: there
is
a grand design in the Universe that favours the development of intelligent
 life''~\cite{8}.
  
Anyway in this paper, in order to revise the Carter's formula, 
we answer Hawking's objection as well. We discuss some assumptions of the
Carter's
model, 
and we stress the importance of the Universe extent both 
in the probability estimation 
of intelligent life  evolution, and 
in the living time of a civilization (Section  2). 
We show that these quantities 
depend not only on the evolution critical steps, but also 
on the number of places where life can arise.

In the light of 
our results,
a stronger formulation of Anthropic
Principle is required, and we give this formulation trying to
conciliate both the Copernican and Anthropic principles, following
the trend started by Gott III~\cite{7b} (section 3). Finally we briefly
discuss the teleological implications of this point of view (section 4).

\section{Carter's formula revisited}

\subsection{Mankind evolution probability on Earth is worse than expected}
Carter's model rests essentially on three steps.

\begin{itemize}
\item The probability that an unlikely evolutionary step,
which is thus ``slow'', compared with the majority of evolutionary steps, 
will happen at time $t$ is
\begin{equation}
 P_i (t) = 1 -  e^{-{t\over \alpha_i}} \simeq  {t\over \alpha_i}
\end{equation}
 where $\alpha_i$ is the timescale for the occurrence of the
 $i$th ``improbable step", with the requirement $\alpha_i >> t_0$, 
where $t_0$ is the biosphere existence length.

\item The $n$ ``improbable steps'' 
are statistically independent,
\begin{equation}
P(t)= \prod^n_{i=1} {t\over \alpha_i}
\end{equation}

\item The conditional probability that mankind evolves at time $t$, 
given that it occurs on before $t_0$ is
\begin{equation}
P'(t) = \left( {t\over t_0} \right)^n 
\end{equation}
\end{itemize}

Using (3), the expectation value ${\overline t} \simeq t_e$ 
for the appearance instant of intelligent life is
\begin{equation}
{\overline t} \equiv \int_0^{t_0} t \, dp' = t_0 {n \over {n+1}}  
\end{equation}
This value implies a strong bound to the time a biosphere 
will continue to evolve in the future
\begin{equation}
t_0 - {\overline t} = {{t_0}  \over {n+1}}  
\end{equation}

This bound has been used to 
exclude the existence of extraterrestrial
intelligent beings: most earth-like planets around $G$ type stars will
be destroyed long before or just after 
intelligent beings have a good chance of evolving. 

Why does this result  not hold on Earth? It is trivially true that humans 
must have evolved much before
life ceased on Earth, 
but {\em reasonable} values of $n$ give {\em unreasonable} 
values of  $t_0 - {\overline t}$, even assuming  
the ``optimistic'' Carter's estimation. \footnote{We do not think, 
as also pointed out by Carter~\cite{3}, that there are $n$ steps in the
Homo Sapiens evolution which are statistically independent,
but  we would rather believe that the events are chained 
in such a way that one step may occur only if  the previous one has happened.
Our opinion is supported by several studies about the evolution (see for
example~\cite{8c}).
If this is true, Carter's evaluation appears to be overestimated 
(This claim is proved in the Appendix I).}

Might Carter's derivation be an argument not only against
the existence of extra terrestrial intelligence, but also against
our own existence? 
Or has Earth  some special properties compared to the rest of the Universe?

\subsection{Intelligent life evolution probability in the Universe 
is better than expected}
Just 
following the Mediocrity Principle we must think that Earth is not a
special 
locus in the Universe but a number $N$ of earth-like planets certainly
exist, and this number is related to the number of galaxies\footnote{
 The Drake's formula~\cite{9} allows a probabilistic estimation
of the technological
civilizations we could find in our Galaxy. 
If we take only some terms of the formula and we consider in the same way
all the other observable galaxies,  we can obtain the probable
number $N$ of earth-like planets in the Universe where an evolution could
have
started (Details are reported in the Appendix II).} 
Could this influence the Carter's estimation?

It is very easy, by using a binomial distribution, to compute the
probability that a number $K$ of civilizations can develop on these $N$
planets,
starting from the hypothesis of the Carter's model. For the sake of 
simplicity we 
consider $\alpha_i \simeq \alpha_j \quad \forall i,j$.

\begin{equation}
P(K civilizations) = \left({N \atop K}\right)
\left( {t\over \alpha} \right)^{nK}
\left[ 1- \left( {t\over \alpha} \right)^n
\right]^{N-K}
\end{equation}

In this case the development probability
of at least one civilization all over the observable 
Universe is
\begin{equation}
P(t)=1-\left[1- 
\left( {t \over  \alpha} \right)^n 
\right]^N
\end{equation}
This evidences that the number $n$ of very improbable steps can be, 
and in fact it is,
balanced by the abundance ($N$) of trials. See for example Figure 1,  
where we report a plot of $P(t)$ as function of $N$, once
fixed $t, \alpha$, and $n$.

This result holds also when we
substitute the general expression 
\begin{equation}
P(t)=1-\left[ 1- \prod^n_{i=1} \left(1- e^{-{t \over \alpha_i}}\right)
\right]^N
\end{equation}
to the previous one.

\subsection{How long does a biosphere remain habitable 
after intelligent life evolution?}
Now we modify the third Carter's step (3), based on
the 
fact that an intelligent species has been produced before $t_0$
on Earth, by conditioning the probability (7) on the evidence
that at least one
civilization has developed all over the Universe.

By applying the Bayes formula we have 
\begin{equation}
P'(t)={{1-\left[1- 
\left( {t \over  \alpha} \right)^n 
\right]^N}\over {1-\left[1- 
\left( {t_0 \over  \alpha} \right)^n 
\right]^N}}
\end{equation}

Then  we 
compute the expected appearance time 
of at least one civilization  all over the observable 
Universe, given that it is found on at least one planet before $t_0$. 

By setting $\gamma_N = {1 \over {1-\left[1- 
\left( {t_0 \over  \alpha} \right)^n 
\right]^N}}$ we have

\begin{equation}
{\overline t} =
\gamma_N \int_0^{t_0} {d \over {dt}}
\left[ 1 -
\left( 1-{ {t^n}\over{\alpha^n}} \right)^N
\right]tdt
\end{equation}
\begin{equation}
= \gamma_N{{Nn}}
\int_0^{t_0}{ t^n \over{\alpha^n}}
\left( 1-{{t^n}\over{\alpha^n}}\right)^{N-1}dt
\end{equation}
By the Newton formula
\begin{equation}
\left(1-{{t^n} \over {\alpha^n}}\right)^{N-1}
= \sum_{k=0}^{N-1} \left({{N-1} \atop k}\right) 
\left({t \over \alpha}\right)^{nk} (-1)^k 
\end{equation}
we have
\begin{equation}
{\overline t} =
\gamma_N Nn \int_0^{t_0}\sum_{k=0}^{N-1} 
\left({{N-1} \atop k}\right) 
\left({t \over \alpha}\right)^{n}  
\left({t \over \alpha}\right)^{nk}(-1)^k dt
\end{equation}
\begin{equation}
= t_0 \gamma_N
Nn \sum_{k=0}^{N-1} \left({{N-1} \atop k}\right) 
{{(-1)^k} \over {n(k+1)+1}}
\left({t_0 \over \alpha}\right)^{n(k+1)} 
\end{equation}
and, when $N=1$, 
we recover the Carter's result (4).

So the time a biosphere will continue to evolve in the future is
\begin{equation}
t_0-{\overline t} =
t_0 \left( 
1 - \gamma_N Nn \sum_{k=0}^{N-1} 
\left({{N-1} \atop k}\right)
{{(-1)^k} \over {n(k+1)+1}}
\left({t_0 \over \alpha}\right)^{n(k+1)} 
\right)
\end{equation}
We can see, in Figure 2, ${\overline t}$ as a function of $N$,
and, in Figure 3, the corresponding behaviour of 
$t_0-{\overline t}$  in terms of $N$ 
(in the plot the values are scaled by $t_0$). 
It is easy to verify that the expected living time of a civilization 
increases with
the number of earth-like planets in the Universe\footnote{For
 computing purposes a useful
approximation of (14) is given by
$${\overline t} \leq t_0 
{{n}\over {n+1}} \gamma_N N
\left({t_0 \over \alpha}\right)^{n}
\left[1-\left({t_0 \over \alpha}\right)^{n}\right]^{N-1}$$}

\section{Mediocrity Anthropic Principle}
So we can answer 
Hawking's objection with our demonstration that the abundance of
creation is {\it necessary} for the life evolution:
the occurrence of intelligent life is 
related, according to the equations (7,9,15),  to 
the enormous number of galaxies. 
 It seems that the constraints on the initial conditions and 
universal constants invoked by
AP 
are not enough to avoid contradictions with the mankind existence
(Carter's formula (5) is an example).
They form just a necessary but not sufficient condition. 

We suggest a stronger 
version of SAP. 
A stronger formulation of SAP must include the existence
of a large number $N$ of earth-like planets such to balance the 
number of improbable steps  
$n$ necessary for evolution. 
It can be formulated in this way: 
{\em ``The Universe (and hence both the fundamental 
parameter on which it depends, {\bf and the amount of places where the
evolution
can take place}) must be such as to admit the creation of
observers at some stage, {\bf  and to assure them a not trivial living
time}''}.
As it rests on the Mediocrity Principle,
we call it {\it Mediocrity Anthropic Principle} (MAP).

\section{A little bit of teleology: classical and quantum finalism}

In our results 
there is no compelling evidence of a
``design'' in the Universe.
We can think that intelligent life is born by
chance thanks to the enormous number of galaxies.

On the other hand a finalist interpretation is still possible.
In fact we want to stress that there are different kinds of finalism. 

 In classical
mechanics one can strike a target with an arrow using suitable initial
conditions: the finalist strategy is the choice of the initial
velocity $\vec{v_0}$.

 In quantum mechanics, if we consider a potential barrier $V$ and we
want to
detect at least one particle with energy $E<V$ on the other side,
 we must use another
strategy. We cannot calibrate the initial conditions so that the winning
behaviour is to shoot a lot of particles through the barrier. When the
game is ruled by probabilistic laws, the abundance of attempts is the
best strategy to follow: this is an example of quantum finalism.

In the cosmological case the equations (7,9,15) suggest that both ways have
to be followed. It seems that the fine tuning of initial conditions and
universal
constants is not enough to assure the birth of Homo Sapiens. 
The extent and abundance of creation in the
Universe could complete the right finalist strategy.

\section*{Acknowledgments}
The authors are grateful to Gaetano Scarpetta and Rossella Sanseverino
 for useful suggestions.

\section*{Appendix I}
In this Appendix we show that Carter's evolution probability (2)
is an overestimation when the evolution steps are chained.

In this framework we divide the timescale in discrete 
intervals of $\tau$ width and suppose that
the occurrence probability of the $i$th ``improbable step" at a
given instant of time $t_i = m_i \tau $, given 
it did not occur in the $m_i-1$
previous instants is
\begin{equation}
{1\over \alpha_i}\left( 1-{1\over \alpha_i}\right)^{m_i-1}
\end{equation}
If we require the occurrence of $n$ events in a fixed series of instants
such
that $0<t_1< t_2<...<t_n$, the probability of the whole sequence is
\begin{equation}
\prod^n_{i=1} {1\over \alpha_i} \left( 1-{1\over
\alpha_i}\right)^{m_i-m_{i-1}-1}
\end{equation}
For the sake of simplicity we can assume that $\alpha_i \simeq \alpha_j $
so that
the
equation (17) becomes:
\begin{equation}
\left({1\over \alpha}\right)^n \left(1-{1\over
\alpha}\right)^{m_n - n}\quad \hbox{where} \quad m_n - n =\sum^n_{i=1}
\left(m_i-m_{i-1}-1\right)\quad \hbox{and}\quad m_0=0
\end{equation}
In this way the probability the Homo Sapiens evolves by a
time
$t = t_n$ is
\begin{equation}
P(t) = {{n \tau}\over t}
\left({{t \over {\tau}}\atop n}\right) \left({1\over
\alpha}\right)^n \left( 1-{1\over
\alpha}\right)^{{t \over \tau}-n} 
< \left({{t \over {\tau}}\atop n}\right) \left({1\over
\alpha}\right)^n \left( 1-{1\over
\alpha}\right)^{{t \over \tau}-n} 
\end{equation}
where the first term takes into account all the possible choices of
$t_1, t_2,...,t_{n-1}$ such that $0<t_1< t_2<...< t_{n-1}<t_n=t$. 
If we assume a great value for $t_0$, we can approximate (19) 
by a Poisson distribution, and neglect $\tau$, by assuming
$\alpha$ timescaled in the same way as $t$. 
\begin{equation}
P(t) = {1\over n!}\left({t\over {\alpha }}\right)^n e^{-{t\over {\alpha}}}
<< \left({t\over {\alpha }}\right)^n \end{equation}

\section*{Appendix II}
Frank Drake conceived an approach, the {\em Drake equation},
to bound the terms involved in
estimating the number of technological civilizations that may exist in 
our galaxy. The Drake equation identifies specific factors which
play a role in the development of such civilizations. 
From this equation we take only the followings:
\begin{itemize}
\item $R_*$, the rate of formation of suitable stars,
\item $f_p$, the fraction of stars with planets (Extra sun system
planets are proved to exist.
A recent example is the one discovered near the star Lalande 21185.),
\item $n_e$, the number of earth-like planets per planetary system,
\item $f_l$ the fraction of those planets where life develops.
\end{itemize} 

These factors have been evaluated on the about
10 billions of known galaxies, to obtain
an estimation $N$ of earth-like planets  in the Universe.
Although there is a questionable estimation of the parameters of this
equation, 
a not-optimistic evaluation 
puts $N$ in the order of $10^{13}$.

\vfill\eject

\end{document}